\documentclass[12pt]{iopart}

\expandafter\let\csname equation*\endcsname\relax
\expandafter\let\csname endequation*\endcsname\relax

\usepackage{amsfonts}
\usepackage{amssymb}
\usepackage{amsmath}

\usepackage{array}
\usepackage{booktabs}
\usepackage{tabu}
\usepackage{dcolumn}
\usepackage{graphicx}% Include figure files
\usepackage{subfigure}
\usepackage{dcolumn}% Align table columns on decimal point

\usepackage{xcolor}

\begin{document}

\title[]{Null geodesic of Schwarzschild AdS with Gaussian matter distribution}

\author{Abolhassan Mohammadi$^1$,N. Doustimotlagh$^2$, Tayeb Golanbari$^1$, Behrooz Malakolkalami$^1$, Shahram Jalalzadeh$^3$}

\address{$^1$Department of Physics, Faculty of Science, University of Kurdistan,  Sanandaj, Iran.}
\address{$^2$Supreme National Defense University, Tehran, Iran.}
\address{$^3$Departmento de Física, Universidade Federal de Pernambuco, Pernambuco, PE 52171-900, Brazil.}
\ead{\mailto{a.mohammadi@uok.ac.ir}; \mailto{doustimotlagh@chmail.ir}; \mailto{t.golanbari@uok.ac.ir}; \mailto{b.malakolkalami@uok.ac.ir}; \mailto{shahram@df.ufpe.br}}
\vspace{10pt}
\begin{indented}
\item[]August 2021
\end{indented}

\begin{abstract}
One of the best ways to understand the gravitation of a massive object is by studying the photon's motion around it. We study the null geodesic of a regular black hole in anti-de Sitter spacetime, including a Gaussian matter distribution. Obtaining the effective potential and possible motions of the photon are discussed for different energy levels. The nature of the effective potential implies that the photon is prevented from reaching the black hole's center. Different types of possible orbits are considered. A photon with negative energy is trapped in a potential hole and has a back and forth motion between two horizons of the metric. However, for specific values of positive energy, the trapped photon still has a back and forth motion; however, it crosses the horizons in every direction. The effective potential has an unstable point outside the horizons, which indicates the possible circular motion of the photon. The closest approach of the photon and the bending angle are also investigated.
\end{abstract}

\vspace{2pc}
\submitto{\CQG}
\maketitle

%%%%%%%%%%%%%%%%%%%%%%%%%%%%%%%%%%%%%%%%%%%%%%%%%%%%%%%%%%%%%%%%%%%%%%%%%%%%
%%%%%%%%%%%%%%%%%%%%%%%%%%%%%%%%%%%%%%%%%%%%%%%%%%%%%%%%%%%%%%%%%%%%%%%%%%%%
%%%%%%%%%%%%%%%%%%%%%%%%%%%%%%%%%%%%%%%%%%%%%%%%%%%%%%%%%%%%%%%%%%%%%%%%%%%%
%%%%%%%%%%%%%%%%%%%%%%%%%%%%%%%%%%%%%%%%%%%%%%%%%%%%%%%%%%%%%%%%%%%%%%%%%%%%
%============  Sec.I (Introduction)  =======================================
%%%%%%%%%%%%%%%%%%%%%%%%%%%%%%%%%%%%%%%%%%%%%%%%%%%%%%%%%%%%%%%%%%%%%%%%%%%%
%%%%%%%%%%%%%%%%%%%%%%%%%%%%%%%%%%%%%%%%%%%%%%%%%%%%%%%%%%%%%%%%%%%%%%%%%%%%
%%%%%%%%%%%%%%%%%%%%%%%%%%%%%%%%%%%%%%%%%%%%%%%%%%%%%%%%%%%%%%%%%%%%%%%%%%%%
%%%%%%%%%%%%%%%%%%%%%%%%%%%%%%%%%%%%%%%%%%%%%%%%%%%%%%%%%%%%%%%%%%%%%%%%%%%%
\section{Introduction}
Black holes are known as a region of spacetime where gravity is so strong that nothing, even light, can escape from it. It is a natural prediction of general relativity that sufficient mass could change spacetime and form a black hole. The black hole boundary is specified by the event horizon in which nothing inside the horizon has a chance to escape. Then, black holes are entirely black, and we could see nothing inside. Nevertheless, the situation is different near the event horizon. The quantum field theory in curved spacetime indicates that it is possible to emit radiation near the event horizon. The point was first pointed out by Steven Hawking \cite{Hawking:1974sw}, who worked out the mathematical argument, and the radiation is known as Hawking radiation. \\
The discovery of Hawking radiation is the first window for quantum gravity and has raised hope for constructing a unifying theory \cite{Nicolini:2008aj}. The quantization of general relativity, so that gravitons carry the gravitational force, ended up in infinities. Then, gravity was considered entirely  classical. Hawking could find a semi-classical argument to support his conjecture about the black hole evaporation \cite{DeWitt:1974na,DeWitt:1975ys,Birrell:1982ix}. However, it was valid only when emitted particles' energy was small compared to the black hole's mass \cite{Nicolini:2008aj}.The black hole evaporates, and its mass decreases, and eventually, at a certain point, the conjecture was failed to be hole because the semi-classical description is broken down. As the decay continues, a quantum theory of gravity is required to explain the process. So far, there are two candidates for quantum gravity: string theory and loop quantum gravity in which both have their own merit and drawbacks \cite{Rovelli:2004tv}. \\
Regarding our understanding of quantum gravity, black holes are assumed to be our best chance \cite{Nicolini:2008aj}.
%For instance, studying mini black hole could release very good information about quantum gravity and very early universe.
Although everything seems to work for understanding the theory of quantum gravity, the problem is that the candidate theories do not explain the final stage of the black hole. A complete quantum description of black hole evaporation is unlikely since the scattering amplitudes and cross sections by perturbative techniques are not computed. The new approach for solving the problem is the Noncommutative Geometry arguments \cite{Woronowicz:1987wr,Connes:1994yd,Connes:1987rs,Connes:1996gi,Landi:1997sh,Madore:2000aq,vanSuijlekom:2015iaa,
GraciaBondia:2001tr,Connes:2000by}. It is widely believed that there should be an uncertainty in the theory of quantum theory to prevent the measuring of position to an accuracy better than Planck length \cite{Nicolini:2008aj}. The statement comes to this conclusion that the theory should satisfy a new form of uncertainty indicating that the positions do not commute, i.e. $[\mathbf{x}_i,\mathbf{x}_j] \neq 0$ (refer to \cite{Nicolini:2008aj} for more information). The interesting point about including the noncommutativity to the theory is that the spacetime  coordinates become noncommutative operators by means of $[\mathbf{x}^\mu,\mathbf{x}^\nu] = i \alpha^{\mu\nu}$
\footnote{The tensor usually is presented by $\theta^{\mu\nu}$. However, the change has been made to avoid any confusion with the coordinate $\theta$ of the metric},
where $\alpha^{\mu\nu}$ is an anti-symmetric matrix with the length squared dimension. \\
Noncommunity modification has appealing results in both quantum field theory and general relativity \cite{Nicolini:2008aj,Khosravi:2006es,Khosravi:2007vh,Khosravi:2007zb,Bina:2007wj,Bina:2010ir,Jalalzadeh:2013zwa,Rashki:2019mde}. It has been concluded that, under such modification, the final stage of black hole evaporation so that the temperature no longer diverges and the curvature reaches a finite value \cite{Nicolini:2008aj}. The modification alters Einstein’s action, which brings out new field equations \cite{Moffat:2000gr,Jevicki:1998rr,Madore:1993br,Chamseddine:1992yx,Chamseddine:2000si,Chamseddine:2000zu,
Calmet:2006iz,Mukherjee:2006nd,Aschieri:2005zs}. Another viewpoint is that the noncommutative effects leaves the Einstein tensor unchanged, and the effects only appear in the matter source \cite{Nicolini:2008aj,Smailagic:2003rp,Smailagic:2003yb,Nicolini:2005vd,Anacleto:2020zfh}. One consequence of the noncommunity effect is that, in flat spacetime, it  removes the point-like structures in favor of smeared objects \cite{Nicolini:2005vd}. In the mathematical view, the work is done by replacing everywhere the position Dirac-delta function with a Gaussian distribution of minimal width $\sqrt{\alpha}$. The coordinate commutator encodes an intrinsic uncertainty, which implies that instead of having a particle mass $M$ localized in a point, it spreads out through a region of linear size $\sqrt{\alpha}$. The resulting mass density of a static, spherically symmetric, smeared, the point-like gravitational source is presented by Eq.\eqref{rhoGD}. Replacing the vacuum with Gaussian matter distribution leads to a new regular metric. The metric has been considered  for different aspects  \cite{Nicolini:2008aj,Nicolini:2005vd,Anacleto:2020zfh,Nicolini:2005gy,Nicolini:2005zi}. \\
The present work's primary goal is to study the geometry of the null geodesics of AdS Schwarzschild with Gaussian matter distribution. The main motivation of the present work is two folds. First, we have the non-singular nature of the metric. Assuming that general relativity is valid up to arbitrarily high energy and curvature, it was proved by Hawking and Penrose that spacetime with a black hole is not geodesically complete, and there will be a singularity. This result was obtained under the dominant energy condition. The singularity, infinite energy, and curvature are nonphysical and undesirable for us. It is widely believed that formulating a quantum theory of gravity will release gravitation from singularities. One approach to formulating quantum gravity is the noncommutative geometry effects, which leads to some uncertainties and minimal length. The second motivation stands on the importance of null-geodesics. The motion of the massless particle is recognized as way to understand the gravitational field around a black hole. Due to this fact, the null geodesics has been the topic of many research works, where different types of black holes have been considered. In some cases, the black holes are assumed to be surrounded by dark energy candidates such as quintessence. \cite{Zhou:2011aa,Fernando:2012ue,Ghaderi:2017ltc,Mahmoodzadeh:2017ejn,Ghaderi:2017yfr,Bautista-Olvera:2019blb,
Chatterjee:2019rym,Bambhaniya:2019pbr,Mondal:2020pop,Nozari:2020tks}.  \\

The paper is organized as follows: In Sec. II, the spacetime we are working on is introduced. We work with the Lagrangian of the black hole in Sec.III. Then, in Sec.IV, the radial motions are studied where the angular momentum is ignored. Null geodesic with angular momentum is considered in Sec. V, where the effective potential, energy levels, classical force, and circular orbits are discussed. The light passes for different energy levels of the photon are investigated in Sec. VI. Finally, in Sec.VII, the closest approach is considered, and we calculate the bending angle of the photon. The results are concluded in Sec.VIII.\\

%=============================================================%
%=============================================================%
%=============================================================%
%=============================================================%
%============== Section 2 =======================================%
%=============================================================%
%=============================================================%
%=============================================================%
%=============================================================%
\section{The spacetime}
\begin{figure}
  \centering
  \includegraphics[width=8cm]{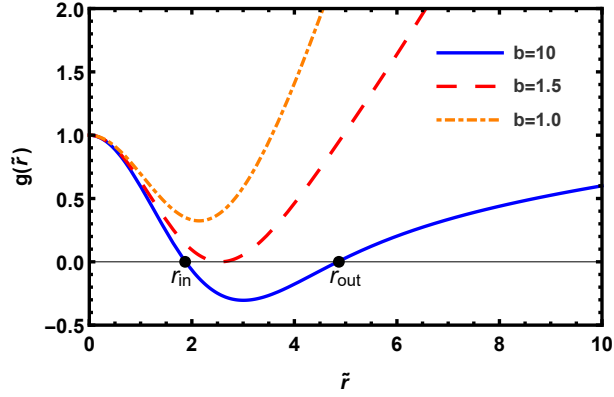}
  \caption{The metric coefficient $g(r)$ versus the radius $\tilde{r}\equiv r/\sqrt{\alpha}$ is plotted for different values of $b$. The other constants are taken as $\tilde{M}\equiv M/\alpha = 2.5$, and $\alpha=0.1$}\label{metricgrAdS}
\end{figure}

\begin{figure}
  \centering
  \includegraphics[width=8cm]{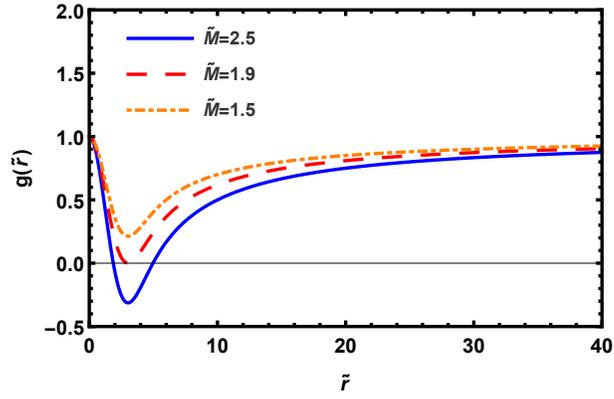}
  \caption{The metric coefficient $g(r)$ (for $\Lambda=0$) versus the radius $\tilde{r}\equiv r/\sqrt{\alpha}$ is plotted for different values of $b$, where $\alpha$ is taken as $\alpha=0.1$.}\label{metricgr}
\end{figure}

Assuming that the spacetime is filled with a negative cosmological constant. Solving the Einstein equation under this condition leads to the regular Schwarzschild anti-de Sitter (AdS) metric, which is static and possesses a spherical symmetry. Adding a matter with Gaussian distribution, as follow \cite{Nicolini:2008aj,Nicolini:2005vd,Nicolini:2011dp,Smailagic:2012cu}
\begin{equation}\label{rhoGD}
  \rho(r) = {M \over \big( 4\pi \alpha \big)^{3/2}} \; e^{-r^2 / 4\alpha},
\end{equation}
where the parameter $\alpha$ indicates the variance of the distribution (i.e., $\sqrt{\alpha} \;$), will change the geometry of spacetime. The presence of such energy density is due to the non-commutative effect, which states that the coordinates do not commute and there is a minimal length; as explained in the introduction. Then, we have no longer a point-like particle, instead, the mass ($M$) of the particle is spread in a region with size $\sqrt{\alpha}$. In mathematical language, it means that the Dirac delta function should be replaced by the Gaussian distribution function. Then, as $\alpha \rightarrow 0$, the non-commutative effects vanish and coordinates commute. In this limit, the Gaussian distribution function in Eq.\eqref{rhoGD} comes to the Dirac delta function \cite{Nicolini:2008aj,Nicolini:2005vd,Nicolini:2011dp,Smailagic:2012cu}. \\
The matter distribution results an energy-momentum tensor $T_{\mu \nu}$ which under the condition $\bigtriangledown_\mu T^{\mu\nu} = 0$ and $g_{00} = -g^{-1}_{11}$ gets the following form
\begin{equation}\label{energymomentum}
  T^0_0 = T^r_r =  - \rho(r), \qquad T^\theta_\theta = T^\phi_\phi = -\rho(r) - {r \over 2} \; {\partial \rho(r) \over \partial r}.
\end{equation}
The energy-momentum tensor describes an anisotropic fluid. Including this source of energy to the Einstein field equations come to a spherically symmetric solution described by
\begin{equation}\label{metric}
  ds^2 = -g(r) dt^2 + g^{-1}(r) dr^2 + r^2 d\Omega^2,
\end{equation}
where $d\Omega = d\theta^2 + sin^2\theta d\phi^2$, and the metric coefficient $g(r)$ is read as
\begin{equation}\label{gr}
  g(r) = 1 + {r^2 \over b^2} - {4 G_N M \over r \sqrt{\pi}} \; \gamma\left( {3 \over 2} , {r^2 \over 4\alpha} \right),
\end{equation}
where $G_N$ is the Newton's constant, which from now on it is taken as $G_N = 1$, and $b$ is the curvature radius of the AdS space. The function $\gamma\left( {3 \over 2} , {r^2 \over 4\alpha} \right)$ is known as the lower incomplete gamma function described as
\begin{equation}
   \gamma\left( n , x \right) = \int_{0}^{x} dt \; t^{n-1} \; e^{-t} .
\end{equation}
Fig.\ref{metricgrAdS} describes the behavior of the function $g(r)$. It is realized that depending on the values of the constants $\alpha$ and $b$, the metric could have two, one, and no horizon, which is plotted in the figure.
The behavior of the metric coefficient $g(r)$ versus the radius $r$ is illustrated in Fig.\ref{metricgrAdS} for different values of $b$. It is realized that as $r$ approaches zero, the coefficient $g(r)$ tends to one.
%%%%%%%%%%%%%%%%%%%%%%%%%%%%%%%%%%%%%%%
%%%%%%%%%%%%%%%%%%%%%%%%%%%%%%%%%%%%%%%
The main difference with the case of no cosmological constant is that, as the radius goes to infinity, the coefficient gets bigger and bigger. However, for the case $\Lambda=0$, as $r \rightarrow \infty$, the coefficient asymptotically approaches to one; Fig.\ref{metricgr}.
%%%%%%%%%%%%%%%%%%%%%%%%%%%%%%%%%%%%%%%
%%%%%%%%%%%%%%%%%%%%%%%%%%%%%%%%%%%%%%%

%=============================================================%
%=============================================================%
%=============================================================%
%=============================================================%
%============== Section 3 ====================================%
%=============================================================%
%=============================================================%
%=============================================================%
%=============================================================%
\section{Null Geodesics}
The Lagrangian of the described black holes, as stated by
\begin{equation}\label{lagrangian}
  \mathcal{L} = {-1 \over 2} \; \left( -g(r) \dot{t}^2 + g^{-1}(r) \dot{r}^2 + r^2 \dot{\theta}^2 + r^2 sin^2\theta \dot{\phi}^2 \right).
\end{equation}
The metric has two killing vector, i.e. $\partial_t$ and $\partial_\phi$, leading to the two constants of motions
\begin{eqnarray}
  g(r) \dot{t} & = & E_n, \label{CoM1} \\
  r^2 sin^2\theta \dot{\phi} & = & L . \label{CoM2}
\end{eqnarray}
As the initial conditions, it is assumed that $\theta = \pi /2$ and $\dot{\theta}=0$. It results in $\ddot{\theta}=0$, which means that $\theta$ stands in $\pi/2$ and undergoes no change. Therefore, the geodesics are describes in a plane at $\pi/2$. By substituting Eqs.\eqref{CoM1} and \eqref{CoM2} into the Lagrangian (\eqref{lagrangian}), one arrives at
\begin{equation}\label{rdotgeneral}
  \dot{r}^2 + g(r) \; \left( {L^2 \over r^2} + \varepsilon \right) = E_n^2 ,
\end{equation}
in which $\varepsilon = 2 \mathcal{L}$ and $\varepsilon = +1$ and $0$ respectively correspond to the massive and massless particle. Comparing Eq.\eqref{rdotgeneral} with the relation $\dot{r}^2 + V_{eff} = E_n^2$, one could defined an effective potential
\begin{equation}\label{potential}
  V_{eff} = g(r) \; \left( {L^2 \over r^2} + \varepsilon \right).
\end{equation}
To consider the circular orbit of the particle, it is required to realize the relation between $\phi$ and $r$. From Eq.\eqref{CoM2}, one finds
\begin{equation}\label{phirgeneral}
  {d\phi \over dr} = {L \over r^2} \; {1 \over \sqrt{E_n^2 - V_e}}.
\end{equation}
For the rest of the work, we concentrate on the null geodesic so that $\varepsilon = 0$. In this case, the effective potential is given by $V_{eff} = L^2 g(r) / r^2$.

%=============================================================%
%=============================================================%
%=============================================================%
%=============================================================%
%=======================  Section 5 ==========================%
%=============================================================%
%=============================================================%
%=============================================================%
%=============================================================%
\section{Radial null geodesics}
There is no angular momentum for the radial geodesics, i.e., $L=0$. Consequently, the effective potential is null, $V_{eff} = 0 $. Besides, from Eqs.\eqref{CoM1} and \eqref{rdotgeneral}, one arrives at
\begin{equation}\label{rdottdot}
  \dot{r} = \pm E_n , \qquad \dot{t} = {E_n \over g(r)}.
\end{equation}
Combining the relation, the time $t$ is extracted as a function of the radial $t$, so that
\begin{equation}\label{tr}
  {dt \over dr} = {\pm 1 \over g(r)},
\end{equation}
and the time is obtained by taking integration, however, analytically solving the integration faces difficulties de to the presence of incomplete gamma function. \\
%\begin{equation}\label{time}
%  t(r) =
%\end{equation}
The proper time $s$ is derived from the relation $\dot{r} = dr / ds = \pm E_n$ by integration as
\begin{equation}\label{propertime}
  s = \pm {r \over E_n} + c ,
\end{equation}
where $c$ is a constant of integration. It is realized that as the particle approaches the horizons, $r \rightarrow r_{in,out}$, the proper time is always a finite value, $s \rightarrow \pm r_{in,out}/ E_n$.

%=============================================================%
%=============================================================%
%=============================================================%
%=============================================================%
%=======================  Section 3 ==========================%
%=============================================================%
%=============================================================%
%=============================================================%
%=============================================================%
\section{Angular momentum for null geodesics}
\begin{figure}[h]
  \centering
  \includegraphics[width=8cm]{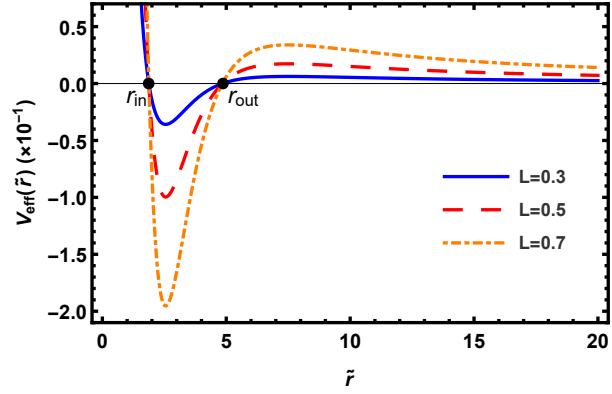}
  \caption{The behavior of the potential versus the radius $\tilde{r}$ for different values of the angular momentum. The other constants are taken as $\tilde{M}\equiv M/\alpha = 2.5$, $\alpha=0.1$, and $b=10$.}\label{potAdS}
\end{figure}

\begin{figure}
  \centering
  \includegraphics[width=8cm]{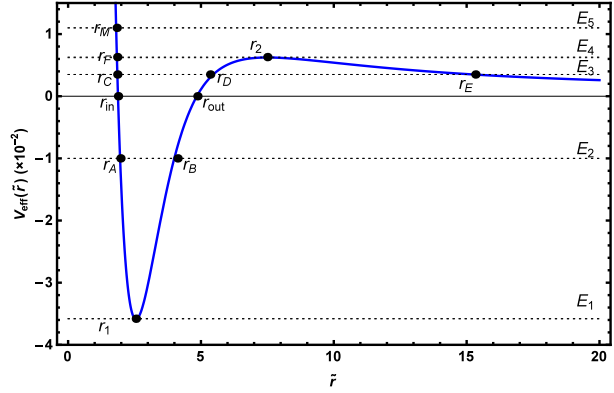}
  \caption{The figure shows the effective potential of the particle and some energy levels. The constants are taken as $\tilde{M}\equiv M/\alpha = 2.5$, $\alpha=0.1$, $b=10$, and $L=0.3$.}\label{potentialenergy}
\end{figure}

\begin{figure}
  \centering
  \includegraphics[width=7cm]{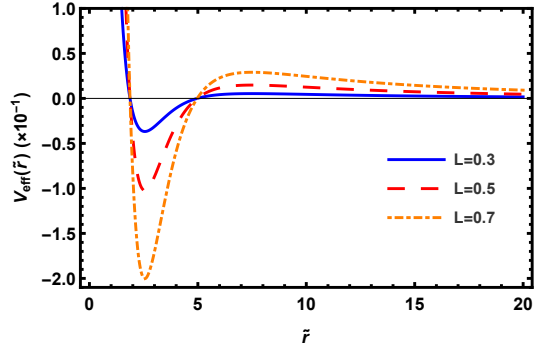}
  \caption{The behavior of the potential versus the radius $r$ for different values of the angular momentum where the negative cosmological constant has been ignored. }\label{pot}
\end{figure}

In the case of $L \neq 0$ the situation will be different. The potential is no longer zero, and it is
\begin{equation}\label{potL}
  V_{eff} = L^2 \; {g(r) \over r^2} .
\end{equation}
Fig.\ref{potAdS} displays the potential of the model versus the radius $r$ for different values of angular momentum. \\
%%%%%%%%%%%%%%%%%%%%%%%%%%%%%%%%%%%%%%%%%
%%%%%%%%%%%%%%%%%%%%%%%%%%%%%%%%%%%%%%%%%
The energy relation $\dot{r}^2 + V_{eff} = E_n$ is utilized to describe the motion of the particle based on its energy level. As an instance, Fig.\ref{potentialenergy} illustrate the effective potential of particle and also there are some energy level. Based on the energy level of the particle, there are different types of motion which are explained as follows: \\
%%%%%%%%%%%%%%%%%%%%%%%%%%%%%%%%%%%%%%%%%
%%%%%%%%%%%%%%%%%%%%%%%%%%%%%%%%%%%%%%%%%
\begin{itemize}
  \item $\mathbf{E=E_1}$: At this energy level, the particle stays in the bottom of the potential hole $r=r_1$, and has a circular orbit. It could not have a radial motion. The shape of the potential implies that the $r=r_1$ point is a stable point. Note that, since $r_{in} < r_A < r_B < r_{out}$, this motion of particle occurs between the two horizons. There is the same situation for all negative energy level.
  \item $\mathbf{E=E_2}$: For these energy levels, the particle is trapped in the potential hole, and it has a radial back and forth motion between the points $r_A$ and $r_B$. In other words, $r_A$ and $r_B$ are the turning points of the particle motion. $\dot{r}$ is zero for $r=r_A$, and $r_B$, but it is not zero in between.
  \item $\mathbf{E=E_3}$: There are two possibilities. The first case is when the particle starts its motion for $r> r_E$. In this case, there is just one turning point $r_E$. The particle approaches the black hole and reaches the point $r_E$. It is then pushed back. The second case occurs when the particle starts its motion in $r_C < r < r_D$. Same as the previous case, the particle is trapped in the potential hole, and it has a radial back and forth motion with two turning points $r_c$ and $r_D$.
  \item $\mathbf{E=E_4}$: With this energy, $\dot{r} = 0$ in $r=r_2$, and the particle has a circular orbit in this radial point. However, this is an unstable point as it is clear from the potential shape. If the particle deviates to the left, the particle falls into the potential hole. The particle crosses the two horizons, but it has a turning point in $r=r_F < r_{in}$. If the particle deviates to the right, it is pushed away from the black hole without any turning point.
  \item $\mathbf{E=E_5}$: There is an unbounded orbit for this energy. The particle moves toward the center of the black hole, cross the two horizons, reaches the turning point $r=r_G$, and then it is pushed away.
\end{itemize}
In general, The particles never reach the center of the black hole. They might even cross the horizons, but they never get to the center. They move toward the center, but for $r<r_1$, they are pushed away, stopped at their turning points. All the orbits are unbounded for energies $E > E_4$. The potential is flat for the considerable radial distance, mainly due to the negative cosmological constant in the metric. The potential reaches   an asymptotical value. If one eliminates the term $\Lambda$, the potential gradually decreases for large radial distance and it tends to zero. It could be realized from Fig.\ref{pot}, which describes the behavior of the potential versus $r$ for different values of the angular momentum.
%%%%%%%%%%%%%%%%%%%%%%%%%%%%%%%%%%%%%%%%%
%%%%%%%%%%%%%%%%%%%%%%%%%%%%%%%%%%%%%%%%%

%=============================================================%
%=============================================================%
%=============================================================%
%=============================================================%
\subsection{Classical force}
\begin{figure}[h]
  \centering
  \includegraphics[width=7cm]{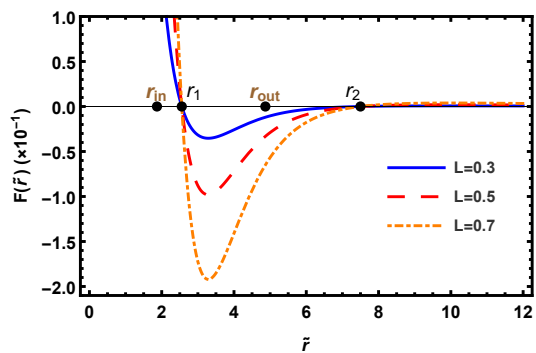}
  \caption{The total imposed on a photon during its motion around the black hole versus the radius $\tilde{r}$ for different values of angular momentum. The other constants are taken as $\tilde{M}\equiv M/\alpha = 2.5$, $\alpha=0.1$, and $b=10$.}\label{totalforceAdS}
\end{figure}

\begin{figure}[h]
  \centering
  \includegraphics[width=8cm]{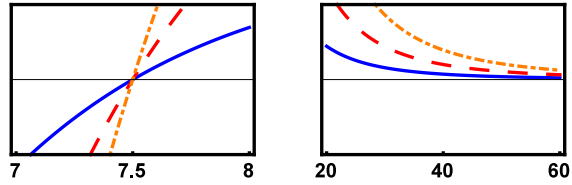}
  \caption{The plot displays the behavior of the total force in the second and third critical point of the radius. It is realized that the total force is zero in the point and it changes sign.}\label{forcezeroAsymAdS}
\end{figure}

The force on the photon is obtained by taking radial differentiation from the effective potential, as
\begin{eqnarray}\label{force}
  F & = & {-1 \over 2} \; {dV_{eff} \over dr}  \\
    & = & L^2 \; \left[ {1 \over r^3 } - {6M \over \sqrt{\pi} \; r^4} \; \gamma\left( {3 \over 2},{r^2 \over 4\alpha} \right) \right. \nonumber \\
     & & \qquad \qquad \left. + {M \over 2 \alpha \sqrt{\pi \; \alpha} \; r} \; e^{-r^2 / 4\alpha} \right]. \nonumber
\end{eqnarray}
The first term on the right-hand side (RHS) is always positive, which implies a positive force that pushes the particle away from the center. This force is high for small $r$ and rapidly decrease by enhancement of $r$. The cosmological constant has no role in the classical force. The cosmological constant's central role is in the potential, which brings an asymptotical flat (non-zero) potential for large $r$. The second and third terms on the RHS are due to the presence of Gaussian matter distribution. The total force that a photon experience is depicted in Fig.\ref{totalforceAdS} versus the radius for different values of angular momentum. The force changes sign two times in $r_1$ and $r_2$.\\
%%%%%%%%%%%%%%%%%%%%%%%%%%%%%%%%%%%%%%%%%
%%%%%%%%%%%%%%%%%%%%%%%%%%%%%%%%%%%%%%%%%%%
For small $r < r_1$, the force is positive and it tends to push out the photon. The point $r_1$ is inside the horizons. The photon feels this preventing force when it is inside the horizons and wants to reach the black hole's center. The force gets larger as the photon comes closer to the center and prevents it from reaching the center.  In $r=r_1$, the force changes sign and become negative for $r_1<r<r_2$. The photon is pulled back to the center (note that the outer horizon is in between). The force is zero again for $r=r_2$, and it is positive for $r>r_2$. Then, for $r>r_2$, the photon feels a positive force that drives it away from the black hole. The plot at the left side of Fig.\ref{forcezeroAsymAdS}  enlarges the force's shape at $r=r_2$ which clearly illustrates that the force becomes zero and changes sign. Then, the force is positive and gets larger in magnitude for a while and again gets weaker and tends to zero, depicted in Fig.\ref{forcezeroAsymAdS}. Zero force for large $r$ was expected from the shape of the potential where it was mentioned that the potential becomes flat for large $r$. Therefore, when the photon is at a large distance from the black hole, it almost feels no force.
%%%%%%%%%%%%%%%%%%%%%%%%%%%%%%%%%%%%%%%%%
%%%%%%%%%%%%%%%%%%%%%%%%%%%%%%%%%%%%%%%%%%%

%=============================================================%
%=============================================================%
%=============================================================%
%=============================================================%
%=======================  Section 4 ==========================%
%=============================================================%
%=============================================================%
%=============================================================%
%=============================================================%
\subsection{Circular orbits}
When the particle has a constant energy, i.e. $E_n = E_c$,  the particle's radius remains unchanged,  $\dot{r} = 0$.  At this point, the radial derivative of the effective potential is zero $dV_{eff}/dr=0$,  and the point is an unstable point. Then, the potential and energy get equal, and the radial coordinate of the particle remains unchanged,  $r=r_c$. \\
Suppose that the constants of the model have been justified  in a way that there are two horizons. Then, there will be two $r_c$ describing circular orbits.  \\
The energy $E_c$ and the angular momentum $L_c$ at the circular orbits are related through the impact parameter $D_c$ at the circular stage. The relation is read as
\begin{equation}\label{EcLc}
  {E_c^2 \over L_c^2} = {g(r_c) \over r_c^2} = {1 \over D_c^2} .
\end{equation}
From Fig.\ref{potentialenergy}, it is realized that there might be two circular orbits occurring in $r= r_1$ and $r_2$ respectively related to the two energy levels $E_c = E_1$, and $E_4$. The first circular orbit stays between the two horizons, which is happening at the stable point of the potential. The second circular orbit is outside of the horizon, which stands on the unstable point.

%=============================================================%
%=============================================================%
%=============================================================%
%=============================================================%
\subsection{Time period of circular orbit}
\begin{figure}[h]
  \centering
  \subfigure[\label{Prtimeperiod}]{\includegraphics[width=7cm]{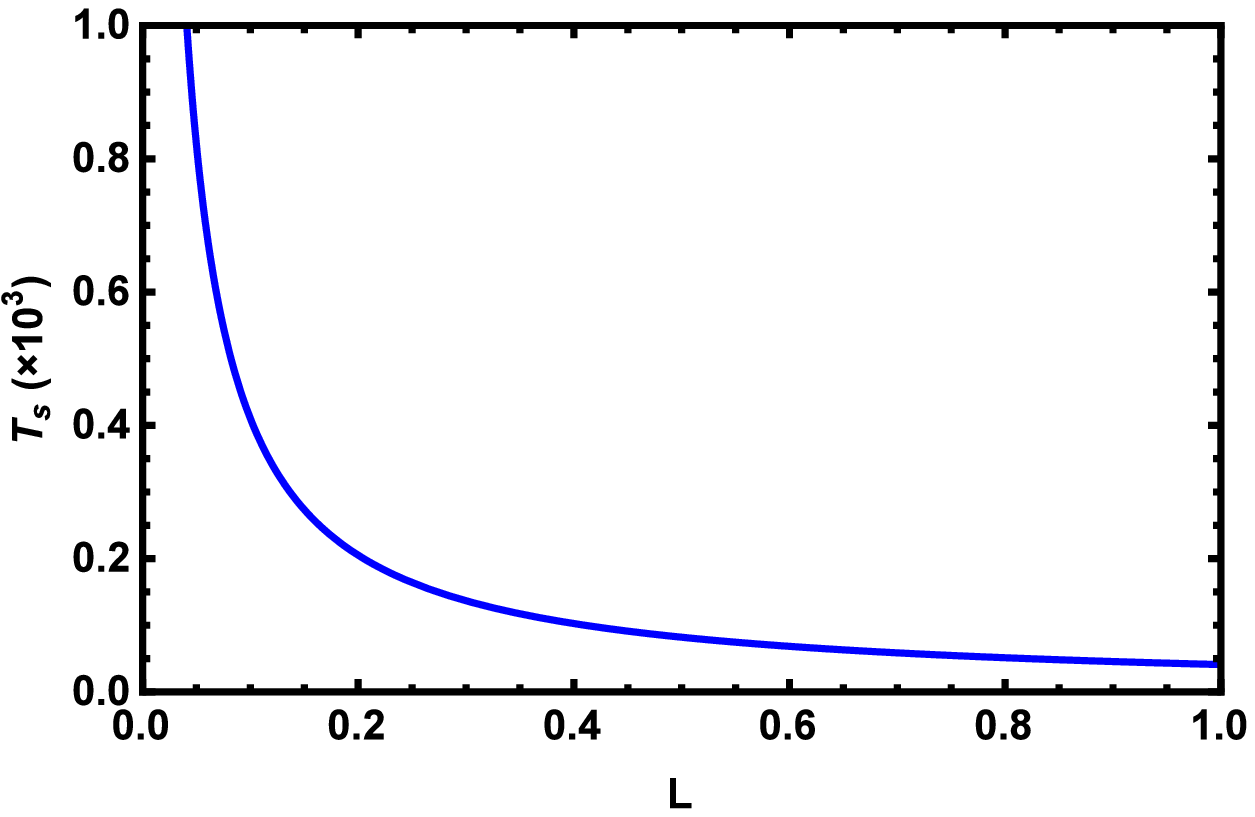}}
  \subfigure[\label{Cotimeperiod}]{\includegraphics[width=7cm]{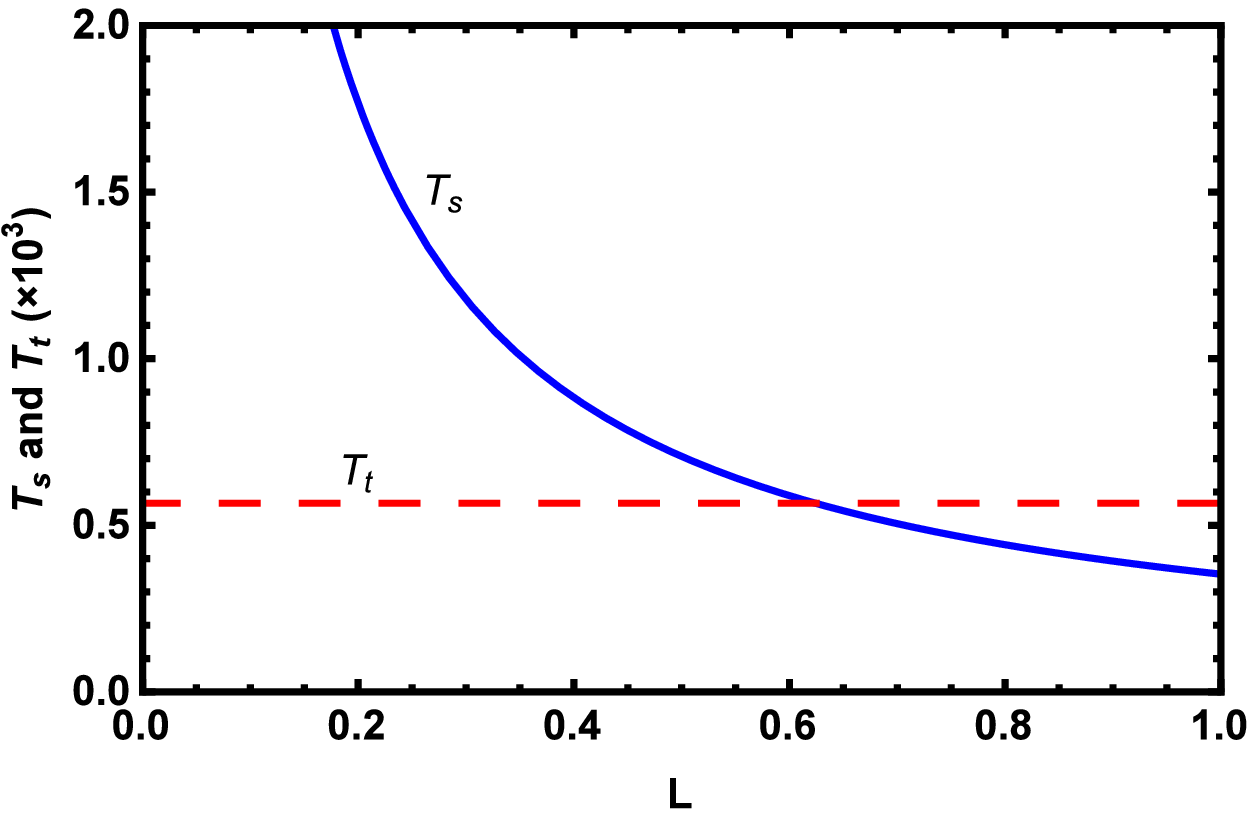}}
  \caption{The proper and physical time period versus angular momentum are plotted for $r_{c2}$ and $r_{c3}$ circular orbits. To plot the quantities, the constants are taken as $\tilde{M}\equiv M/\alpha = 2.5$, $\alpha=0.1$, and $b=10$.}\label{timeperiod}
\end{figure}

The circular orbit's period is derived via the Eq.\eqref{CoM1}  in terms of proper time and physical time. Since the radial $r_c$  for the circular motion is constant, the proper time is obtained as
\begin{equation}\label{propertimeperiod}
  T_s = {2\pi r_c^2 \over L_c},
\end{equation}
and in terms of the physical time, the period is obtained as
\begin{equation}\label{physicaltimeperiod}
  T_t = {2\pi r_c \over \sqrt{g(r_c)}}.
\end{equation}
Both periods are depicted in two plots regarding the two circular orbits in $r=r_{1}$ and $r_{2}$ in Fig.\ref{timeperiod}.
%%%%%%%%%%%%%%%%%%%%%%%%%%%%%%%%%%%%%%%%%
%%%%%%%%%%%%%%%%%%%%%%%%%%%%%%%%%%%%%%%
At $r=r_{1}$, the circular orbit is inside the two horizons and is hidden from the outside observer. Then there is no access to the particle for the outsider observer to measure the physical time. Due to this fact, Fig.\ref{timeperiod}a only exhibits the proper period of the first circular orbit versus the angular momentum. The proper time period increases by decreasing the angular momentum. \\
For the second circular orbit, which stays outside of the horizons, the two time period is displayed in Fig.\ref{timeperiod}b.  For
larger angular momentum values, the physical period is higher. However, the proper period is higher for smaller angular momentum. Note that $T_t$ is independent of the angular momentum.

%=============================================================%
%=============================================================%
%=============================================================%
%=============================================================%
\subsection{Unstable circular orbit}
The Lyapunov exponent $\lambda$ \cite{Cardoso:2008bp}, which is defined as
\begin{equation}\label{Lexp}
  \lambda = \sqrt{-V''_{eff}(r_c) \over 2 \dot{t}^2(r_c)},
\end{equation}
is measure the instability of the unstable circular null geodesics. For the present case, the parameter $\lambda$ is obtained as
\begin{equation}\label{lambda}
  \lambda = \sqrt{-V''_{eff}(r_c) r_c^2 g(r_c) \over 2 L^2}.
\end{equation}
Then, from Eq.\eqref{potential}, it is realized that the instability parameter $\lambda$ does not depend on the angular momentum. There are two circular orbits, one stable and the other unstable, which stays outside the horizon. The instability parameter for the outsider orbit is about $\lambda = 0.83$.
%%%%%%%%%%%%%%%%%%%%%%%%%%%%%%%%%%%%%%%%%%%
%\begin{figure}[h]
%  \centering
%  \includegraphics[width=7cm]{lambdaAdS.eps}
%  \caption{The parameter $\lambda$ versus the angular momentum is plotted for $r=r_{c2}$ and $r_{c3}$}\label{unstableAdS}
%\end{figure}
%%%%%%%%%%%%%%%%%%%%%%%%%%%%%%%%%%%%%%%%%%%

%=============================================================%
%=============================================================%
%=============================================================%
%=============================================================%
%=============================================================%
%=============================================================%
%=============================================================%
%=============================================================%
\section{light path}
\begin{figure}[h]
  \centering
  \subfigure[]{\includegraphics[width=7cm]{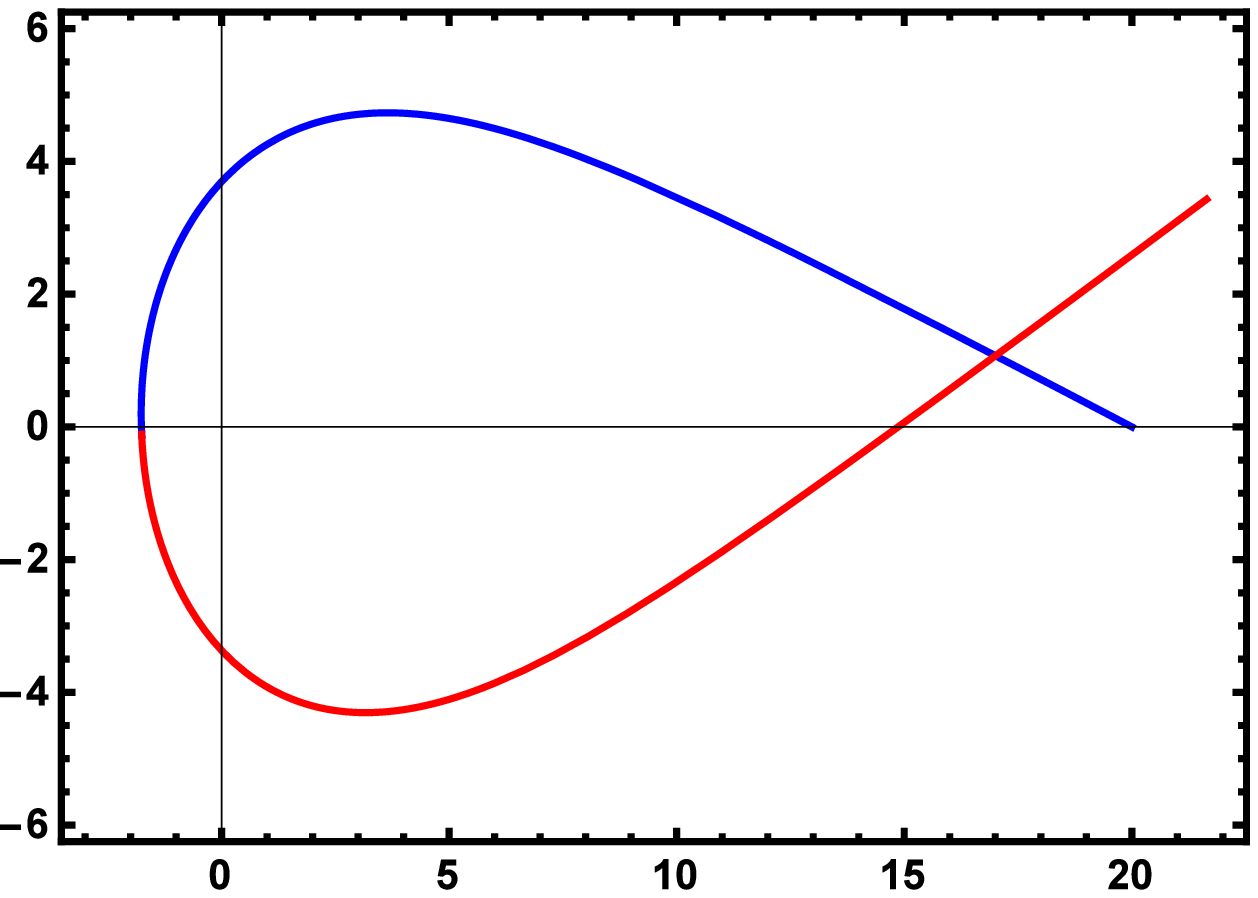}\label{AhighOrbitAdS}}
  \subfigure[]{\includegraphics[width=7cm]{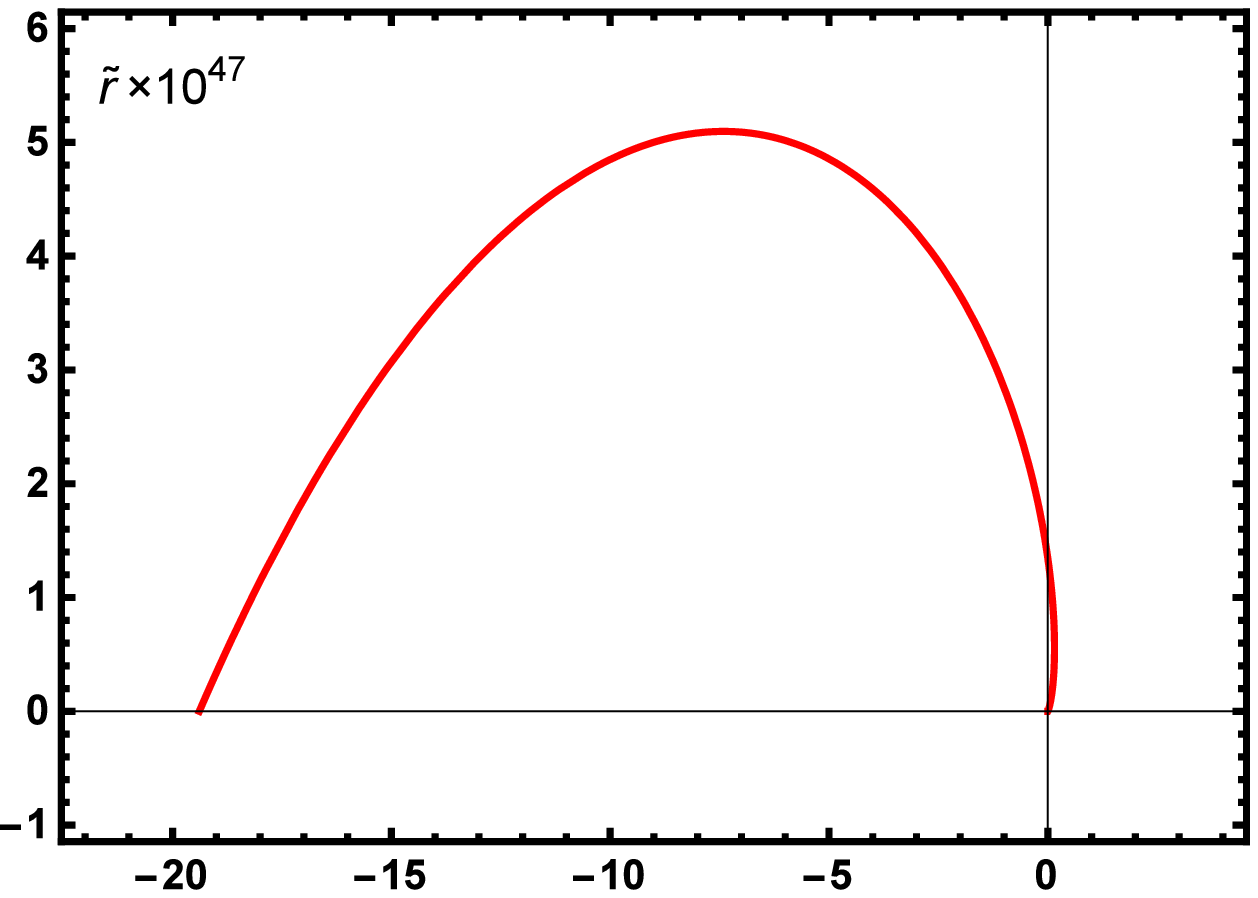}\label{BhighOrbitAdS}}
  \caption{The unbounded orbit of photon with high energy is presented. The incoming and outgoing photon are respectively illustrated by blue and red lines. The figure is a polar plot, i.e. $r-\phi$, which is drawn in $x-y$ coordinate (the $\theta=\pi/2$ plane). The photon is assumed to start from $(r,\phi)=(20,0)$, approaches the black hole, crosses two horizons, reaches the turning point and changes its direction. Then, the photon recedes from black hole and its distance rapidly grows, as presented in (b).}\label{highOrbitAdS}
\end{figure}

\begin{figure}[h]
  \centering
  \subfigure[]{\includegraphics[width=7cm]{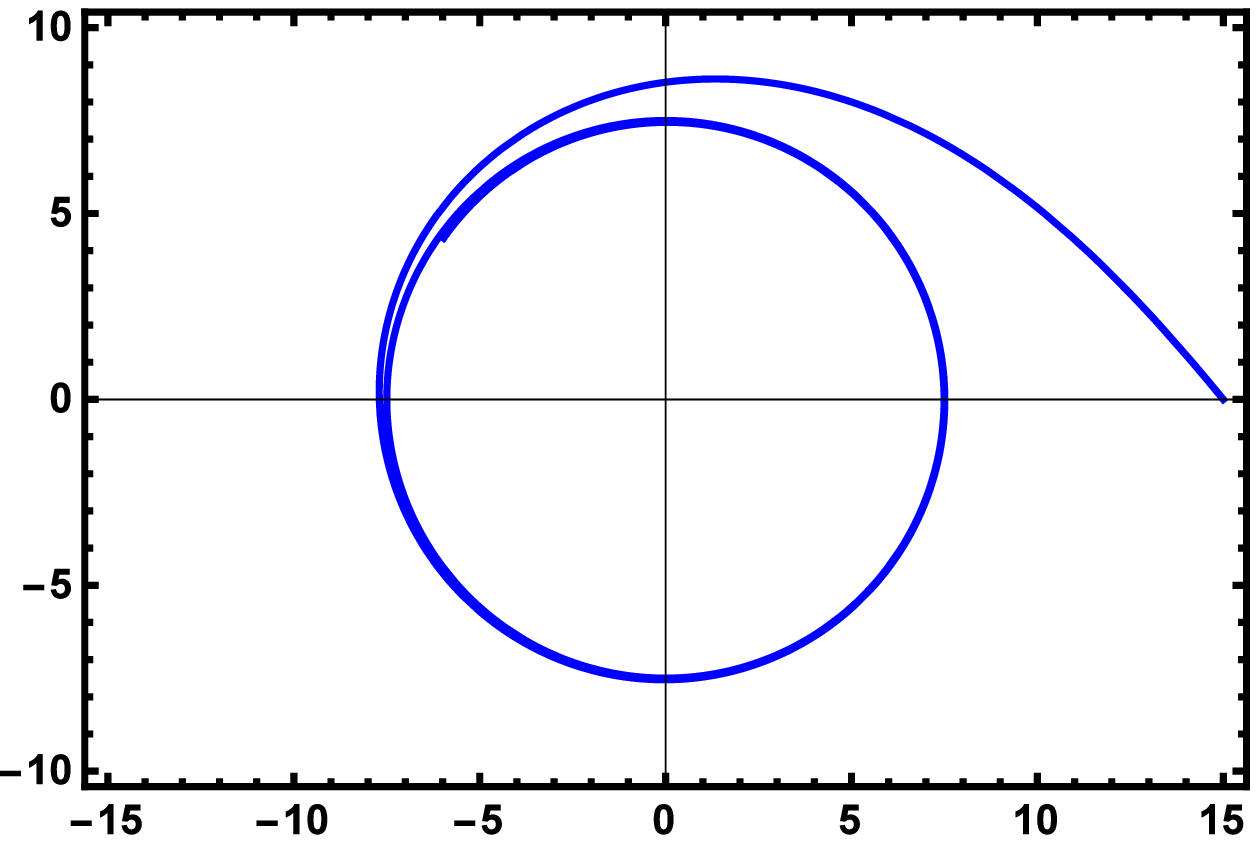}\label{ACorbitAdS}}
  \subfigure[]{\includegraphics[width=7cm]{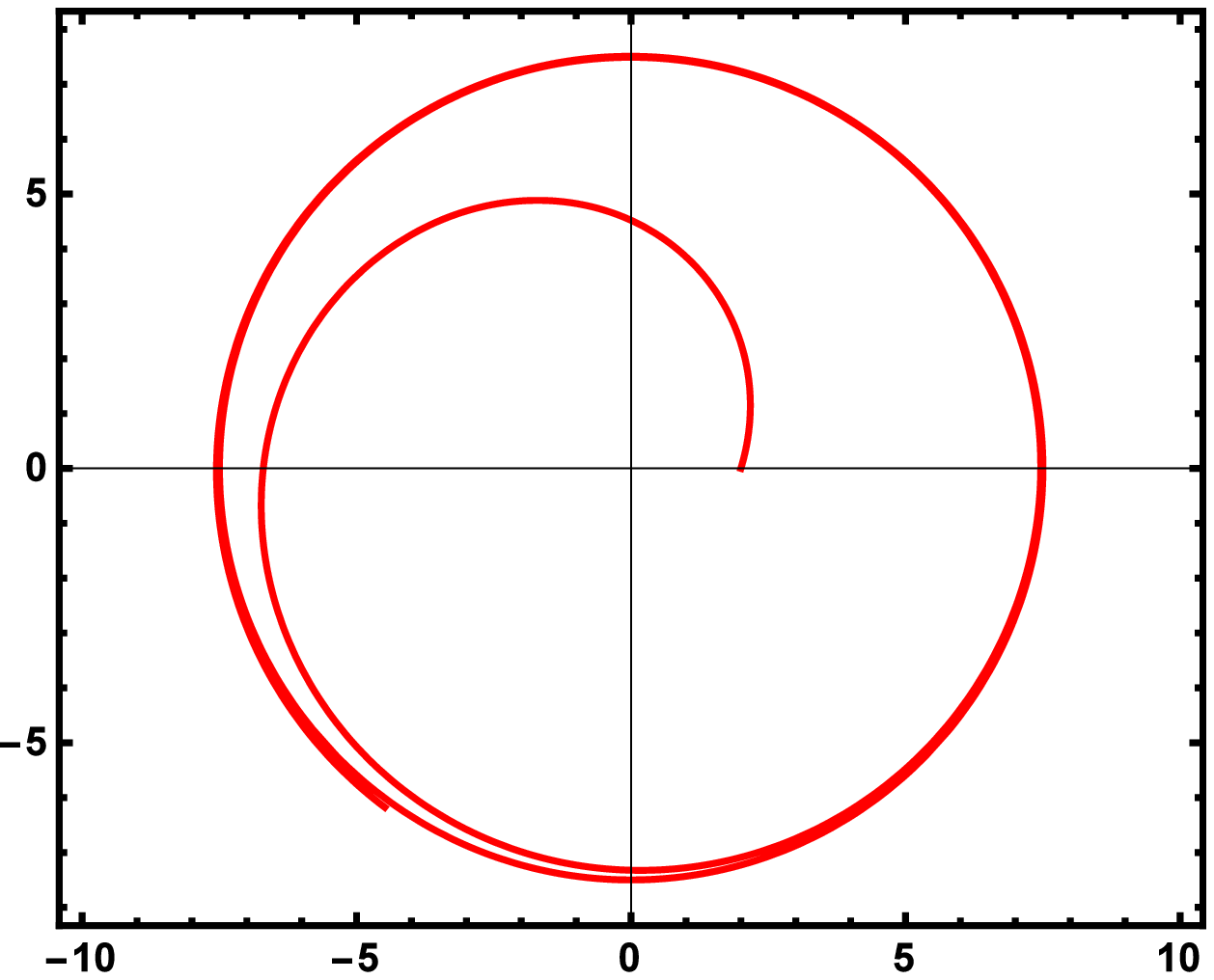}\label{BCorbitAdS}}
  \caption{The blue (red) line present the orbit of incoming (outgoing) photon. The photon approaches (recedes from) black hole. The effective potential increases and at the point $r=r_2$ the effective potential and photon energy are equal. The point is an extremum point of potential (an unstable point). Then, the radial distant of the photon remains unchanged and photon start a circular orbit. }\label{CorbitAdS}
\end{figure}

\begin{figure}[h]
  \centering
  \subfigure[]{\includegraphics[width=7cm]{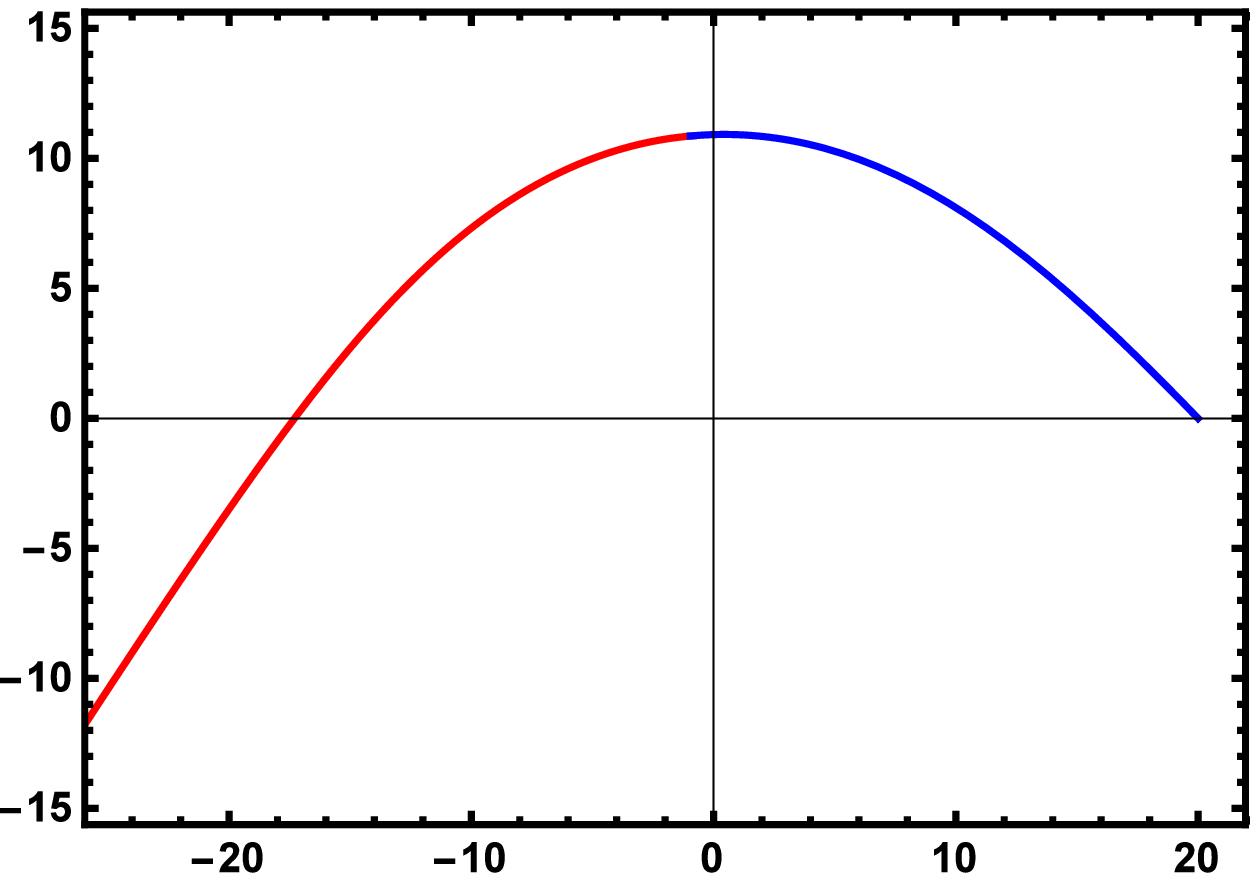}\label{AMediumOrbitAdS}}
  \subfigure[]{\includegraphics[width=7cm]{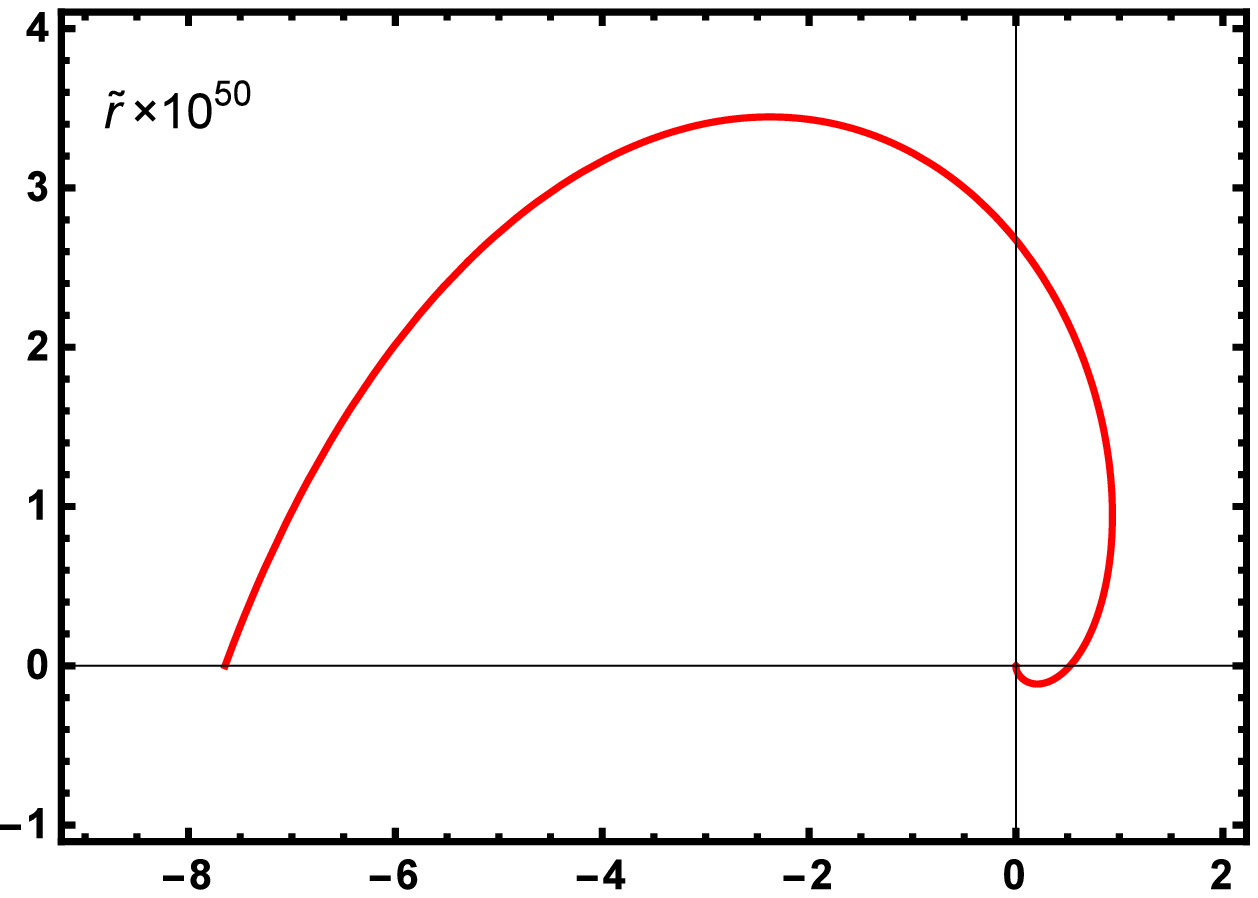}\label{BMediumOrbitAdS}}
  \caption{The plot represents the orbit of photon with medium energy, $L^2/b^2 < E < E_c$, away from the outer horizon. The blue (red) line presents the incoming (outgoing) photon. The photon comes to the black hole, encounters the turning point out of the $r_{out}$ horizon. Then, it turns back and its radial distance $r$ rapidly increases; as presented in (b). }\label{MediumOrbitAdS}
\end{figure}

The effective potential has been considered, and we could describe the possible path of light based on its energy level. Then, the photon's force was investigated, and it was specified whether the force is attractive or repulsive. Now, we are going to consider the path of light and its orbit in more detail. \\
First, we consider an unbounded photon, namely a high energy photon that crosses two horizons and encounters the potential bond in the turning point $r_F$. The photon then comes back and recedes from the black hole. The path is depicted in Fig.\ref{highOrbitAdS}.
%%%%%%%%%%%%%%%%%%%%%%%%%%%%%%%%%%%%%%%%%
%%%%%%%%%%%%%%%%%%%%%%%%%%%%%%%%%%%%%%%%%%%
The blue line describes he incoming photon with start point $(r,\phi)=(20,0)$, depicted in Fig.\ref{AhighOrbitAdS}. The photon approach to the center and reaches the turning point (end of the blue line). Then, the particle changes direction and recedes from the black hole which is shown with the red line in Fig.\ref{AhighOrbitAdS}. The return movement is very fast in which the particle keeps a huge distance by $\phi = 3\pi$ which is illustrated in Fig.\ref{BhighOrbitAdS}. \\
A photon path with energy $E=E_c$ is portrayed in Fig.\ref{CorbitAdS}. The incoming photon is plotted in Fig.\ref{ACorbitAdS} where the start point is assumed to be $(r,\phi)=(15,0)$. The photon approaches the black hole and its potential increases constantly. At point $r=r_2$, the effective potential of the photon equals to the its energy, i.e. $V_{eff}(r=r_2)=E_c$. The radial point $r_2$ is an equilibrium point, specified in Fig.\ref{potentialenergy}, and the particle continues its movement as  a circular orbit. However, it should be noted that the point is an unstable equilibrium point.
%%%%%%%%%%%%%%%%%%%%%%%%%%%%%%%%%%%%%%%%%
%%%%%%%%%%%%%%%%%%%%%%%%%%%%%%%%%%%%%%%%%%%
Fig.\ref{BCorbitAdS} describes an outgoing photon with the same energy level, $E=E_c$. The photon recedes from the center of black hole, crosses the two horizons and reach the point $r=r_2$. At this point, the potential energy of the photon reaches to its energy and the photon has no radial movement any more. Then, it continues its motion as a circular orbit.\\
For a photon with energy lower the energy $E_c$, there are two situations. First, suppose that the photon is away from the outer horizon of the black hole. In this case, for an incoming photo, it approaches to the black hole, encounters to the turning point (like the $r_E$ point in Fig.\ref{potentialenergy}) where $V_{eff} = E$. Then, the photon changes its direction and recedes from the black hole. The light path for this case is presented in Fig.\ref{MediumOrbitAdS}. The blue line in Fig.\ref{AMediumOrbitAdS} describes the incoming photon which is assumed to start its motion at $(r,\phi)=(20,0)$. The radial distance decreases, and the photon comes close to the black hole. It reaches the turning point, which is the end of the blue line. The motion is then reversed, and the photon recedes, which is shown with the red line. Fig.\ref{BMediumOrbitAdS} describes the returning movement where it is realized that the photon gets a considerable distance from the black hole for lateral angle $2\pi$.
%%%%%%%%%%%%%%%%%%%%%%%%%%%%%%%%%%%%%%%%%
%%%%%%%%%%%%%%%%%%%%%%%%%%%%%%%%%%%%%%%%%%%
The blue line in Fig.\ref{AMediumOrbitAdS} describes the incoming photon which is assumed to start its motion at $(r,\phi)=(20,0)$. The radial distance decreases, and the photon comes close to the black hole. It reaches the turning point, which is the end of the blue line. The motion is then reversed, and the photon recedes, which is shown with the red line. Fig.\ref{BMediumOrbitAdS} describes the returning movement where it is realized that the photon gets a considerable distance from the black hole for lateral angle  $2\pi$. \\
The last path that is considered is for a photon in the potential hole.
%%%%%%%%%%%%%%%%%%%%%%%%%%%%%%%%%%%%%%%%%
\begin{figure}[h]
  \centering
  \includegraphics[width=8cm]{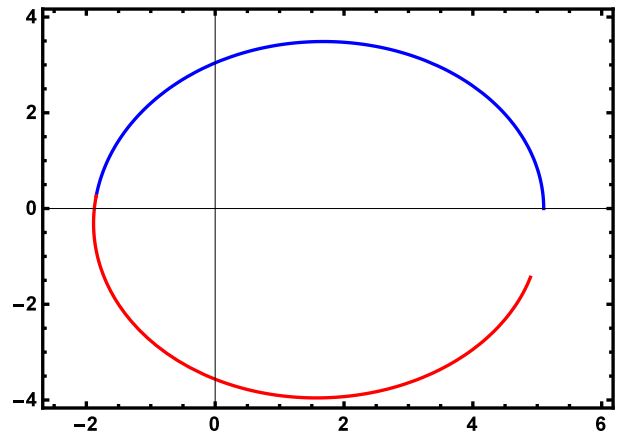}
  \caption{The figure illustrates the orbit of a photon trapped in potential hole. The energy of photon is not enough to escape from the hole and it only has a back and forth motion. The polar plot of the orbit is drawn in figure, which is pictured in $x-y$ coordinate. The blue (red) line presents the incoming (outgoing) photon. The photon starts at $(r,\phi)=(5,0)$, approaches the center, faces the turning point, which is a point on the wall of the potential hole. Then, it turns back, and again reaches the second turning point which also stands on the wall of the potential hole. The photon do not have enough energy to leave the hole and it continues its back and forth motion.   }\label{inhole}
\end{figure}
%%%%%%%%%%%%%%%%%%%%%%%%%%%%%%%%%%%%%%%%%%%
A photon with this energy could not escape from the hole and it has a back and forth motion, like a photon with energy $E_3$ in Fig.\ref{potentialenergy} that has two turning point, $r_c$ and $r_D$. The photon start point is assumed to be $(r,\phi)=(5,0)$. It starts its orbit, and the radial distance decrease until it reaches the turning point; the path is depicted with the blue line in Fig.\ref{inhole}. Then, it returns, takes an orbit with increasing radial distance, and shows with the red line until it reaches the other turning point. This path repeats, and the photon could never escape because it does not have enough energy.

%=============================================================%
%=============================================================%
%=============================================================%
%=============================================================%
%============== Section 2 =======================================%
%=============================================================%
%=============================================================%
%=============================================================%
%=============================================================%
\section{Gravitational Lensing}
One of the main application of null geodesics are gravitational lensing and finding the bending of light. The first step to this matter is to take an unbounded orbit with energy $E=E_0$. The light approached the black hole, and it reaches the minimum distance, $r=r_0$, and then recedes from the black hole. We need to find how photons deviate from their original path due to the black hole's gravity. In this regard, it is required first to find the closest approach $r_0$.

%=============================================================%
%=============================================================%
%=============================================================%
%=============================================================%
\subsection{Closest approach $r_0$}
\begin{figure}[h]
  \centering
  \includegraphics[width=8cm]{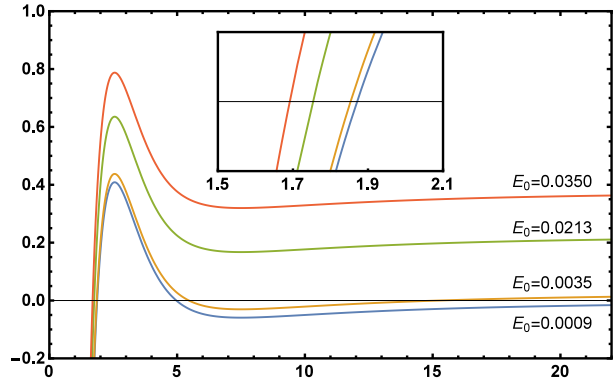}
  \caption{The plot displays the behavior of the function $\chi(r)$ versus the radius $r$ for different values of energy. the point $\chi(r_0)=0$ is clear from the figure. It is realized that for selected energy, there could be found one, two, or three $r_0$ which mainly depends on the magnitude of energy. The constants are taken as $\tilde{M}\equiv M/\alpha = 2.5$, $\alpha=0.1$, $b=10$, and $L=0.3$.}\label{closestr0}
\end{figure}

An unbounded orbit of light has been picked out. So, the photon approaches the black hole, reaches a minimum distance $r_0$ (close approach), and then keeps out. The orbit occurs in $\theta = \pi /2$ plane. In general, the distance $r$ of light changes with the angle $\phi$. Therefore, to find the closest approach, and one needs to solve $dr / d\phi = 0$. From Eqs.\eqref{CoM1}, \eqref{potential} and \eqref{rdotgeneral}, it is realized that
\begin{equation}\label{rphi}
  \left( {1 \over r^2} \; {dr \over d\phi} \right)^2 = {E_n \over L^2} - {1 \over r^2} - {1 \over b^2} +
  {4M \over \sqrt{\pi} \; r} \; \gamma\left( {3 \over 2} , {r^2 \over 4\alpha} \right) \equiv \chi(r) .
\end{equation}
The energy level has a vital role in determining the closest distance $r_0$. From the potential, Eq.\eqref{potential} and Fig.\ref{potAdS}, it is realized that there is no unbounded orbit for negative energy. As a matter of fact, the potential asymptotically approaches to a constant value $L^2 / b^2$ for large radial distance. Then, the minimum energy which should be picked is $L^2 / b^2$. Fig.\ref{closestr0} illustrates the function $\chi(r)$ versus the radius $r$ for different values of energy. It is exhibited that there are one, two, or three values of $r$ leading to $\chi=0$ depending of the magnitude on energy. \\
%%%%%%%%%%%%%%%%%%%%%%%%%%%%%%%%%%%%%%%%%
%%%%%%%%%%%%%%%%%%%%%%%%%%%%%%%%%%%%%%%%%%%
For larger energy, The photon approaches the black hole for more considerable energy, crosses two horizons, gets to the closest distance $r_0$, and then recedes from the black hole. Therefore, there is only one closest distance $r_0$ for the photon with energy $E > V_{eff}(r_2)$ . \\

%=============================================================%
%=============================================================%
%=============================================================%
%=============================================================%
\subsection{Bending angle}
\begin{figure}[h]
  \centering
  \includegraphics[width=7cm]{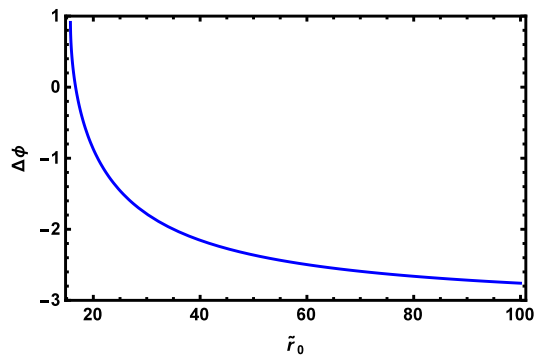}
  \caption{The plots describe the bending angle ($\Delta\phi$) versus the closest approach $r_0$. The higher energy of photon indicates by smaller $r_0$. The constants are taken as $\tilde{M}\equiv M/\alpha = 2.5$, $\alpha=0.1$, and $b=10$. }\label{angle}
\end{figure}

The unbounded photon could be divided into two types. The first corresponds to the photons that have only one turning point, which stays inside the horizons. The second type includes the photon with energy  $L^2/b^2 < E < E_c$ which has three turning points, and only one of them is out of horizons. The other two turning points stand on the potential hole wall, which is related to the trapped photon. Here, we are only interested in unbounded photons. At the turning points, we have $\chi(r)=0$. \\
For static, spherically symmetric metric, with general form of line element
\begin{equation}
  ds^2 = A(r) dt^2 - B(r) dr^2 - D(r) r^2 \Big( d\theta^2 + \sin^2\theta d\phi^2 \Big) ,
\end{equation}
the bending angle is obtained as \cite{Amore:2006xp}
\begin{equation}\label{bendingangleintegral}
  \Delta\phi = 2 \int_{r_0}^{\infty} \; { \sqrt{B(r) \over D(r)} \over \sqrt{ \left( {r \over r_0} \right)^2 {D(r) \over D(r_0)} \; {A(r_0) \over A(r)} - 1 } } \; {dr \over r} \; - \; \pi ,
\end{equation}
where $r_0$ indicates the closest approach \cite{Amore:2006xp}. Comparing the above line element with Eq.\eqref{metric}, it is realized that there are $A(r)=-g(r)$, $B(r)=-g^{-1}(r)$, and $D(r)=-1$, leading to bending angle
\begin{equation}\label{bendingangleintegralour}
  \Delta\phi = 2 \int_{r_0}^{\infty} \; {\sqrt{g(r)} \over \sqrt{ \left( {r \over r_0} \right)^2 \; {g(r) \over g(r_0)} - 1 } } \; {dr \over r} - \pi .
\end{equation}
The integral could not be solved analytically due to the incomplete gamma function. However, it could be solved numerically.
%Solving the integral numerically gives the results presented in Table.\ref{tableAngle}.
%%%%%%%%%%%%%%%%%%%%%%%%%%%%%%%%%%%%%%%%%%%%%%%%%%%%%
%\begin{table}[h]
%  \centering
%  \begin{tabular}{p{1cm}p{1cm}p{1cm}}
%    \hline
%    % after \\: \hline or \cline{col1-col2} \cline{col3-col4} ...
%    $E_0$ & $r_0$ & $\Delta\phi$ \\[0.1cm]
%    \hline
%    $0.0350$ & $1.692$ & $2.823$ \\[0.2cm]
%    $0.0213$ & $1.754$ & $3.533$ \\[0.2cm]
%    $0.0035$ & $15.74$ & $0.945$ \\[0.1cm]
%    \hline
%  \end{tabular}
%  \caption{The bending angle of photon with different energy.}\label{tableAngle}
%\end{table}
%%%%%%%%%%%%%%%%%%%%%%%%%%%%%%%%%%%%%%%%%%%%%%%%%%%%%%
Fig.\ref{angle} presents the bending angle versus the closest approach $r_0$, in which the smaller $r_0$ corresponds to the photon's higher energy. The bending angle is positive for smaller $r_0$ and decreases by enhancement of $r_0$.
%%%%%%%%%%%%%%%%%%%%%%%%%%%%%%%%%%%%%%%%%%%
%%%%%%%%%%%%%%%%%%%%%%%%%%%%%%%%%%%%%%%%%%%
The bending angle asymptotically approaches a constant value that is not zero. The main reason for this result is that the metric is not asymptotically flat. For large radial distance, the AdS term in the metric dominates and causes such behavior.

%=============================================================%
%=============================================================%
%=============================================================%
%=============================================================%
%=======================  Section 5 ==========================%
%=============================================================%
%=============================================================%
%=============================================================%
%=============================================================%
\section{Conclusion}
Including the Noncommutative Geometry arguments in general relativity has been the topic of many research projects. The effect of noncommunity could be considered in two aspects: it could modify the Einstein field equation or it leads to a modified energy density so that removing the point-like structure is one of the noncommunity effects. \\
Solving the Einstein field equations in AdS spacetime, including a Gaussian matter distribution, leads to a metric that describes a regular black hole. In the presented work, the null geodesics, as a tool for understanding a massive source's gravity, were discussed. Through the Lagrangian of the black hole, the effective potential was extracted, which indicates that based on the metric's constants, the described black hole could have two, one, or no horizon. The work was dedicated to the case of possessing two horizons. Based on the effective potential, in a general view, photon's possible motions were considered based on the energy levels. \\
The classical force on the photon and also the possible circular orbits were discussed. There would be two circular orbits, one at the bottom of the effective potential hole, which is a stable point of the effective potential. This orbit would occur between the horizons. Moreover, the second circular orbit is at  $r=r_2$ which is an unstable point of the effective potential. The physical and proper period was estimated. There is no physical time for the first circular orbit between the horizons, and only the proper time is measurable. However, for the second circular orbit, both periods could be found. It was realized that the proper time is smaller than the physical period for small angular momentum. As the angular momentum increases, the proper period will enhance as well, and it could be even larger than the physical period. \\
Next, the path of the photon was discussed in more detail. It was explained that for negative energy, the photon is trapped in the potential hole, and it has a back and forth motion between two horizons, in which the turning points are inside the horizons. For positive energies,  $E_n > L^2 / b^2$,  there are more options. For intermediate energies, i.e., $L^2/b^2 < E_n < E_c$,  the photon could either be still trapped in the potential hole or have an unbounded orbit to infinity. If the photon is trapped in the potential hole, it has a back and forth motion and crosses the horizons in each direction, and changes the direction as it reaches the turning points. A photon with energy $E_c$  has a circular orbit at $r_2$  which is an unstable point of the effective potential. The higher energy of photon leads to an unbounded orbit. However, no matter how much we increase the photon's energy, it never reaches the center of the black hole due to the nature of the potential. The effective potential rapidly grows up as $r$ approaches zero. \\
The closest approach of photon and the bending angle was the last topic considered. The bending angle was obtained positive for small $r_0$ and reduced by growing $r_0$, and asymptotically approaches to a non-zero constant value. Due to the AdS nature of the metric, the AdS term is the dominant term of the metric for large radial distance.

\ack
The work of A.M. has been supported financially by "Vice Chancellorship of Research and Technology, University of Kurdistan" under research Project No.99/11/19063. The work of T. G. has been supported financially by "Vice Chancellorship of Research and Technology, University of Kurdistan" under research Project No.99/11/19305.
%=============================================================%
%=============================================================%
%=============================================================%
%=============================================================%
%=============================================================%
%=============================================================%
%=============================================================%
%=============================================================%

\section*{References}
\bibliography{AdSchwGMD}

\end{document}